# An axiomatic approach to Einstein's boxes

Thomas V Marcella[a]

*Department of Physics and Applied Physics, University of Massachusetts-Lowell, Lowell, Massachusetts, 01854*

The fallacies inherent in the Einstein's Boxes thought experiment are made evident by taking an axiomatic approach to quantum mechanics while ignoring notions not supported by the postulates or by experimental observation. We emphasize that the postulates contain everything needed to completely describe a quantum experiment. We discuss the non-classical nature of both the state vector and the experiment that it represents. Einstein's boxes is then described by the formalism alone. We see that it is no different from any other experiment in which a two-state observable is measured.

## I. INTRODUCTION

Quantum mechanics is a statistical theory about observables and their measurement. It is a theory with modest goals: to predict only the possible results of a measurement and the probability distribution of those results. The postulates are few and the mathematical formalism is straightforward. A course in linear algebra contains most of the mathematics required.[1]

Nevertheless, quantum mechanics is considered by many to be not only difficult, but also downright absurd when compared with classical theory. Textbooks and popular articles alike seem to delight in the most bizarre of interpretations. But, in this presentation, we discuss the nature of quantum mechanics from the mathematical formalism alone. We do not acknowledge any interpretation, other than the Max Born probability interpretation of the state vector. It is the only interpretation that has near universal acceptance, and we embrace it as a postulate.

Prompted by Fuchs and Peres,[2] we take quantum mechanics at face value, never straying too far from the given postulates. We do not speculate about those things not addressed in the formalism or not supported by experimental evidence.

Further, we presume that there is no classical explanation of quantum phenomena.

We begin in Section II by stating the postulates as simply as possible.[3] This idealistic approach provides a complete description of the quantum experiment as defined in section III.

Although we identify the constituent parts of the experimental apparatus, a quantum experiment is defined as a single entity. We describe its non-classical nature and we assert that the only results of the experiment are the measured values of a specified observable. We stress that there is no experiment without a measurement result.

In Section IV we discuss the meaning of the state vector as given by the postulates. We define the quantum state as a vector obtained from the preparation apparatus and that it, especially, does not describe the behavior of the quantum particle. We emphasize that there is no evidence that the state vector has any meaning beyond what is contained in the postulates.

Within this rigid framework, we describe the thought experiment known as 'Einstein's Boxes', the traditional discussion of which, along with a comment, was given in a recent issue of this journal.[4,5] The description given here is straightforward and no different from any other experiment involving a two-state observable. We avoid the bewilderment found in the usual discussion by stressing that the ball and boxes moving in 3-space is not equivalent to the state vector in Hilbert space.

## II. THE POSTULATES OF QUANTUM MECHANICS

**Postulate 1. Definition of observable.**

The observables of quantum mechanics are Hermitian operators that possess a complete set of eigenvectors.

**Postulate 2. Measurement of an observable.**

The only possible results of a measurement are the eigenvalues of the measured observable.

**Postulate 3. Definition of quantum state.**

The quantum state is determined by the preparation procedure. It is represented by the state vector

$$|\psi\rangle = c_k|a_k\rangle = |a_k\rangle\langle a_k|\psi\rangle, \tag{1}$$

where the basis vectors $|a_k\rangle$ are the orthonormal eigenvectors of the measured observable $\hat{A}$. The expansion coefficients $c_k = \langle a_k|\psi\rangle$ are probability amplitudes.

**Postulate 4. The Born probability interpretation**

Given the experiment described by state vector $|\psi\rangle$, the probability that a measurement of observable $\hat{A}$ yields the value $a_k$ is

$$P_k = |c_k|^2 = |\langle a_k | \psi \rangle|^2. \tag{2}$$

**Postulate 5. Time evolution of the state vector**

The quantum state vector evolves in time according to Schrödinger's time-dependent equation

$$i\hbar \frac{\partial}{\partial t} |\psi\rangle = \hat{H} |\psi\rangle, \tag{3}$$

where $\hat{H} = \hat{p}^2/2m + \hat{V}$ is the Hamiltonian operator.

## III. THE QUANTUM EXPERIMENT

A quantum experiment consists of a particle, a preparation apparatus, and a measurement result.[6] The different elements of the experimental apparatus, including the measuring device, make up a whole with no classical analog. Although the components are all classical objects there is no classical explanation of "how the experiment works".

The experiment requires a performed measurement[7] of a specified observable. The pre-measurement and post-measurement experiment are undefined; there is no "before" or "after".

Repeating the experiment does not guarantee the same measurement result; the outcome of the experiment is indeterminate. There is absolutely no way to determine the experimental result prior to measurement. This indeterminate nature is inherent in quantum phenomena. Yet, the experiment reveals nothing about any underlying reality that might account for this uncertainty.

But, repeating the same experiment a large number of times provides us with:

a) all the possible values for the physical quantity being measured and

b) the statistical distribution of those measured values.

The totality of measured results is the eigenvalue spectrum of the observable being measured while the statistical distribution of those results is given by $|\langle a_k | \psi \rangle|^2$.

The statistical distribution $|\langle a_k | \psi \rangle|^2$ is unique for a given experiment. As such, it can be used to identify the experimental configuration, as is done in so called "interaction-free measurements".[8,9]

If the apparatus is modified, or if a different quantity is measured, the new results have no relevancy to the original experiment; different experiments yield different results that are mutually exclusive. The only exception is for commuting observables measured with the same preparation apparatus. But, generally, quantities not measured in the current experiment have no meaning, even though they may have been determined in earlier measurements. This is most evident in the failure of EPR-like experiments[10,11] and Bell's theorem,[12,13] each of which attempts to correlate data from different experiments when there is no such correlation.

Even when the apparatus is changed at the last possible moment, as in "delayed choice experiments",[14] the statistical distribution of the results still corresponds to the apparatus in place at the time of the measurement.

We have no idea how the particle is affected, if at all, as it passes through the apparatus. The quantum experiment yields only the value obtained in the measurement. As such, it does not reveal anything about pre-measurement particle behavior. As Wheeler has said, in discussing a photon experiment, "---------we have no right to say what the photon is doing in all its long course from point of entry to point of detection".[15]

Any interaction of the particle with the preparation apparatus is obviously non-classical. For example, a particle changes direction in passing through a slit apparently without the benefit of a deflecting force. The uncertainty in its momentum brought about by the slits is sufficient to scatter the particle in any direction.[16]

Further, there is no known mechanism that explains how the particle becomes encoded with the information that is manifested in the probability distribution of the measured results.

A measurement requires the detection of the particle. After detection, the particle often becomes lost in the detecting material. And in extreme cases, such as photon detection via the photoelectric effect, the particle no longer exists. In such cases it is impossible to know the state of the particle after the measurement. No matter. Quantum mechanics does not describe particle behavior and the state of the particle after the measurement is of no concern.

The particle is a real object that is always detected as a localized entity. A single particle event provides no evidence of any wave behavior. In particular, interference effects cannot be observed with the detection of a single particle. It is the statistical distribution of many particles, detected at different locations at different times, that is identified as an interference pattern.[17]

Since interference with particles can occur in quantum experiments, the classical maxim, "Waves exhibit interference, particles do not." no longer applies.

## IV. ABOUT THE MEANING OF THE STATE VECTOR

The experimental apparatus in place at the instant the measurement is made defines the state preparation that determines the state vector $|\psi\rangle$. If the experiment is changed in any way, at any time, then the state vector corresponding to the new apparatus must be used in our calculations.

The measured observable determines the actual mathematical form taken by the state vector. It is the measured observable that determines the representation of $|\psi\rangle$ in a complex, linear vector space (Hilbert space); the basis vectors in Hilbert space are the complete set of eigenvectors $|a_k\rangle$ of the measured observable $\hat{A}$.

There is no pre-measurement, or post-measurement, state vector. There is only one state vector $|\psi\rangle$ corresponding to the preparation procedure. The postulates simply tell us that if the purpose of the experiment is to measure observable $\hat{A}$, then we write $|\psi\rangle$ in terms of the eigenvectors of $\hat{A}$.

A state vector is either an eigenvector of the observable or a superposition of its eigenvectors. If we know, *in principle*, the value of observable $\hat{A}$, then we can predict the outcome of the measurement and the state vector is an eigenstate.

But usually, the state vector is an indeterminate superposition state $|\psi\rangle = \sum |a_k\rangle\langle a_k|\psi\rangle$, as defined in postulate 3. The superposition state is a mathematical construct that allows us to calculate the probabilities $P_k = |\langle a_k|\psi\rangle|^2$ when there is more than one possible result of a measurement.

Quantum mechanics does not provide us with the value of observable $\hat{A}$ for such a state and there is no way, *even in principle*, for us to predict the outcome of a measurement. Such indeterminacy is the fundamental characteristic of a superposition state. The human observer has no effect on the quantum state and observer ignorance is not sufficient for the state vector to be indeterminate.

The superposition state $|\psi\rangle$ also enables us to determine other probability-related information. For example, we can calculate the expectation value $\langle \hat{A} \rangle = \sum P_k a_k = \langle\psi|\hat{A}|\psi\rangle$ and the uncertainty $\Delta\hat{A} = \sqrt{\langle \hat{A}^2 \rangle - \langle \hat{A} \rangle^2}$. In a superposition state the observable $\hat{A}$ is indeterminate with an uncertainty $\Delta\hat{A} \neq 0$.

When the observable has a continuous spectrum its eigenfunctions define a complex linear function space of infinite dimensions. In particular, the position representation defines a linear function space in which the state function $\psi(x) = \langle x|\psi\rangle$ is an inner product and, according to the Born postulate, it is a probability amplitude. Calling $\psi(x)$ a "wavefunction" conjures up all our prejudices about classical waves that are real solutions of the classical wave equation. It is less perplexing if we call $\psi(x)$ the "state function".

As a solution of Schrödinger's equation

$$i\hbar \frac{\partial}{\partial t} \psi(x,t) = \hat{H}\psi(x,t) \qquad (4)$$

the state function $\psi(x,t)$ is necessarily complex. As such, $\psi(x,t)$ has no representation in space-time, as do classical (real) wave functions. The formalism does not support the contention that state functions are objective entities that propagate in space-time along with real waves and particles.

Further, there is no empirical evidence that the state function interacts with the measuring device, or that it "collapses" as a result of the measurement process; the state function is not observable and it cannot be measured. Heisenberg,[18] among others, has emphasized that probability amplitudes (wavefunctions) are not physical waves propagating in space-time. And since locality is a property of space-time, state functions in a configuration space present no violation of relativity theory.

We must not confuse the real particle in space-time with the complex state vector in Hilbert space. We emphasize that the state vector is associated with the experimental apparatus, not the particle.

N. G. van Kampen has warned us thus:

"Whoever endows $\psi$ with more meaning than is needed for computing observable phenomena is responsible for the consequences……."[19]

## V.  THE QUANTUM DESCRIPTION OF EINSTEIN'S BOXES

Einstein's boxes is a quantum thought experiment consisting of a ball and two boxes, a preparation apparatus, and a measuring device that counts the number of balls in each box. The ball is a real object that interacts with the measuring device. It is always found in one of the boxes.

Since there is only one ball, the quantity being measured is a two state observable with eigenvalues 1 and 0. The possible results of the experiment are:

Result A:   1 particle in box 1, 0 particles in box 2.
Result B:   0 particles in box 1, 1 particle in box 2.

The corresponding eigenvectors are $|\psi\rangle_A = |n_1 n_2\rangle_A = |10\rangle$ for result A and $|\psi\rangle_B = |n_1 n_2\rangle_B = |01\rangle$ for result B, where $n_1$ is the number of balls in box 1 and $n_2$ is the number of balls in box 2. These eigenvectors are orthonormal:

$$_A\langle\psi|\psi\rangle_B = \delta_{AB}. \qquad (5)$$

The orthogonality condition $\langle 10|01\rangle = 0$ prevents us from ever finding the ball in both boxes.

Because the ball is a conserved entity the result for box 1 is entangled with the result for box 2. If we find the ball in box 1, then we know with certainty that box 2 is empty, even without looking; $n_1 = 1$ always means that $n_2 = 0$. The entangled eigenstate $|n_1 n_2\rangle$ is a single entity corresponding to a single result. It cannot be separated into the two individual state vectors, $|n_1\rangle$ for the ball in box 1 and $|n_2\rangle$ for the ball in box 2.

We assume that the preparation procedure is such that it is absolutely impossible, *even in principle,* to predict which box the ball is in; the state vector $|\psi\rangle$ is indeterminate. Further, if the preparation procedure does not favor one box over the other, then

$$|\psi\rangle = \frac{1}{\sqrt{2}}\left[|10\rangle + e^{i\phi}|01\rangle\right] \qquad (6)$$

Although we cannot predict the outcome of an individual measurement, the state vector does allow us to determine the probability distribution of the results from many measurements:

The probability of obtaining result A is $|\langle 10|\psi\rangle|^2 = \frac{1}{2}$, and the probability of obtaining result B is $|\langle 01|\psi\rangle|^2 = \frac{1}{2}$.

The superposition state $|\psi\rangle$ does not describe the behavior of the ball, or the boxes. Rather, $|\psi\rangle$ describes the preparation procedure from which it is obtained. The formalism provides no way for us to answer such questions as, "What box is the ball in?" or "Is it in both boxes?" or "Is it half in one box and half in the other?" Without a measurement, there is no experiment and quantum mechanics is silent. And, since there is no classical explanation, such questions are meaningless.

We are told that the preparation procedure involves splitting a box containing the ball into the two boxes used in the experiment. So be it. But, splitting the original box into two boxes in 3-space is not equivalent to "splitting" the entangled eigenvector $|n_1 n_2\rangle$ into two separate eigenvectors, $|n_1\rangle$ and $|n_2\rangle$. Transporting one box to Paris and the other to Tokyo prior to measurement has no effect on the outcome and serves no useful purpose, other than to cause confusion.

The boxes are two separate and independent systems that move in a local way in 3-space. But, the entangled state vector consists of a single vector defined in a Hilbert space. The experiment is not described by two spatially separated eigenvectors, one in box 1 and the other in box 2. In fact, there is no evidence, of any kind, that the eigenvectors are contained in the boxes.

This concludes our discussion of Einstein's boxes.

If, on the other hand, there is another experiment for which we do know that the ball is in one of the two boxes, but we just don't know which one, then the different preparation apparatus defines a different state vector:

knowing the ball is in box 1, $|\psi\rangle = |\psi\rangle_A = |10\rangle$.

knowing the ball is in box 2, $|\psi\rangle = |\psi\rangle_B = |01\rangle$.

Here, the preparation procedure always yields an eigenstate; the new state vector $|\psi\rangle$ is not indeterminate. When we find the ball in one box or the other we are observing a pre-existing condition, as in a classical measurement. If repeated experiments yield a statistical distribution of the results, then this is a consequence of classical ignorance, not quantum uncertainty. In such a case, the statistical distribution is obtained from a statistical mixture of the eigenstates $|10\rangle$ and $|01\rangle$. If we could eliminate the ignorance of the observer, then we could predict the outcome of such an experiment.

A statistical mixture of eigenstates is obviously not the same as a superposition state, and in many cases we can experimentally see the difference. For example, in some experiments, a superposition (indeterminate) state leads to interference, while a statistical mixture does not.

The quantum description of the original Einstein's boxes is not incomplete just because this new experiment is described by a statistical mixture. Rather, it simply means that we are doing a different experiment, the results of which cannot be used to provide any information about Einstein's boxes.

## VI. CONCLUDING REMARKS

Many will find this presentation to be too restrictive. Yet, the formalism alone suggests that quantum mechanics is only about the results of measurements - and nothing else. Of course, it is the "and nothing else" that is most bothersome. Many refuse to accept the possibility that there is nothing more to quantum mechanics than what is contained in the postulates.

Bell[20] argued that quantum mechanics should describe more than idealized experiments isolated from physical reality. Likewise, Max Jammer has said, "------- a formalism, even if complete and logically consistent, is not yet a physical theory".[21] Yet, we need only the postulates to completely describe everything that is revealed to us in the quantum experiment. If there is a physical reality that gives rise to quantum probabilities, we know nothing of it.

Quantum theory is silent on the issue of physical reality. As such, it neither supports nor rejects any physical interpretation. But, every attempt at an interpretation entails such a grotesque distortion of physical reality that it is questionable whether there is any advantage to be gained from such pursuits. It does seem that if we want the mechanical universe of classical physics to remain intact then we must be willing to abandon questions of interpretation.

# ACKNOWLEDGEMENT

I wish to thank Daniel Styer for reading the original manuscript. His comments and helpful suggestions are gratefully acknowledged.